\documentclass[review]{elsarticle}
\usepackage{tikz, graphicx}
\usepackage{textpos}
\usepackage{tikz}
\usepackage{ctable}
\usepackage{graphics}
\usetikzlibrary{quantikz}
\usepackage{lineno,hyperref}
\modulolinenumbers[5]

\begin{document}
	
	\begin{frontmatter}
		
		\title{A Four-Party Quantum Secret-Sharing Scheme based on Grover's Search Algorithm}

		\author[mymainaddress]{Deepa Rathi}
		\ead{km.deepa@mt.iitr.ac.in}
		
		\author[secondadress]{Farhan Musanna}
		\ead{farhanmusanna@bitmesra.ac.in}
		
		\author[mymainaddress]{Sanjeev Kumar\corref{mycorrespondingauthor}}
		\cortext[mycorrespondingauthor]{Corresponding author}
		\ead{sanjeev.kumar@ma.iitr.ac.in}
		
		\address[mymainaddress]{Department of Mathematics, Indian Institute of Technology Roorkee, Roorkee, India}
		\address[secondadress]{Department of Mathematics, Birla Institute of Technology Mesra, Ranchi, India}
		
		
		\begin{abstract}
			The work presents an amalgam of quantum search algorithm (QSA) and quantum secret sharing (QSS). The proposed QSS scheme utilizes Grover's three-particle quantum state. In this scheme, the dealer prepares an encoded state by encoding the classical information as a marked state and shares the states' qubits between three participants. The participants combine their qubits and find the marked state as a measurement result of the three-qubit state. The security analysis shows the scheme is stringent against malicious participants or eavesdroppers. In comparison to the existing schemes, our protocol fairs pretty well and has a high encoding capacity. The simulation analysis is done on the cloud platform IBM-QE thereby showing the practical feasibility of the scheme.
		\end{abstract}
		
		
	\end{frontmatter}
	
	
	\section{Introduction}
	\label{intro}
	The purpose of a crypto-system is to protect confidential data over computer network communication in the presence of attackers. The way of sharing secret information among many parties is known as secret sharing in cryptography. Quantum cryptography, an amalgam of quantum mechanics and classical cryptography, can achieve unconditioned security. The security of quantum cryptography relies on the fundamental laws of quantum mechanics instead of complex mathematical problems involved in classical cryptography. In the prospect of quantum cryptography, many branches have been established, such as quantum key distribution (QKD)\cite{1,2,3}, quantum secure direct communication (QSDC)\cite{f,4,5}, quantum network\cite{r,s,t}, identification systems\cite{6}, quantum digital signature\cite{7,8,9}, quantum image encryption\cite{A,B,C}. Similarly, quantum secret sharing (QSS) is one of the most important research problem in secure communication.
	\newline
	The first classical secret sharing scheme was given by Adi Shamir in his seminal paper\cite{10}  ``How to share a secret" in 1979. He used a polynomial to show how to divide a secret into $n$ shares such that at least $k$ of those shares can be used to regenerate the secret, but less than $k$ shares devise no secret. The quantum counterpart of the classical secret-sharing was firstly proposed by Hillery et al.\cite{11}. They considered the entanglement property of the three-particle and four-particle Greenberger-Horne-Zeilinger (GHZ) states to create the shares and detect any participant's spying. Subsequently, Cleve et al.\cite{12} investigated the existence of quantum $(m,n)$-threshold scheme with the constraints as $n \leq 2m$. They also show the conditions for which threshold schemes must distribute the information. Karlsson et al.\cite{13} implemented a $(m,n)$-threshold QSS scheme based on the multi-particle entanglement and entanglement swapping of Bell state, and prevent their scheme from being an eavesdropper by using the non-orthogonal entangled states. Gottesman\cite{14} showed that the only condition on the existence of QSS schemes is the no-cloning theorem and monotonicity. Including this fact, they prove some results about QSS schemes with general access structures. The work of Guo et al.\cite{15} presents a secret sharing scheme independent of the entangled state. Later, Xiao et al.\cite{16} generalized the Hillery et al.\cite{11} QSS scheme into an arbitrary number of parties via introducing the two efficient techniques: favored-measuring-basis and the measuring-basis-encrypted from quantum key distribution (QKD), which increases the efficiency of their scheme asymptotically. Some other QSS schemes with entanglement swapping\cite{17}, Einstein-Podolsky-Rosen (EPR) pairs\cite{18}, product states\cite{19}, and multiparty to multiparty\cite{20} have been reported in the literature to ensure provably secure schemes.
	\newline
	Apart from the above work, several other researchers proposed novel schemes in this direction. Markham and Sanders\cite{21} consider a classical and quantum secret sharing for higher dimensional stabilizer states using graph state formalism. Jia et al.\cite{22} proposed a novel idea for a dynamic quantum secret sharing scheme in which they share classical and quantum information based on a starlike cluster state with $n$ two-qubit arms. The specific feature of their scheme was the addition and deletion of participants during the sharing of classical and quantum information, which makes it more appropriate for practical applications. Yang et al.\cite{23} used the quantum Fourier transform for devising a secret sharing. Mitra et al.\cite{24} adopted a game theory's rationality concept to derive a novel rational quantum secret sharing scheme based on fairness, correctness, and strict Nash equilibrium. Musanna and Kumar\cite{p} harnessed the properties of Bell state sequential measurements to design a novel QSS scheme, which is also applicable in quantum image sharing by using the novel enhanced quantum representation (NEQR). The novelty of their scheme is lay on the dependence of pure quantum mechanical properties of photons. Musanna and Kumar\cite{q} proposed another novel QSS scheme by using GHZ product state to share the secret and Bell measurements to reveal the secret. Some QSS schemes based on the quantum Fourier transform are investigated in Ref.\cite{25,26,27}. Recently, Jiang et al.\cite{28} proposed a verifiable QSS scheme by using the eight-particle entangled state. They analyze the security of their scheme against the internal and external attackers via quantum decoy-state.\\
	Quantum search algorithm (QSA) is a crucial research topic in quantum theory. The QSA is known for the development of Grover's algorithm\cite{29}, which aim is to find a search item in an unclassified database with $N$ particles in $O(\sqrt{N})$ steps. Instead, Grover's search algorithm searching the target item in an unclassified database can be much faster than the best-known classical search algorithm, especially when $N$ is large. In recent years, many researchers adopted Grover's search algorithm to design cryptography protocol like QSS protocol\cite{30,31,32}, quantum direct communication protocol (QDC)\cite{33,34}, quantum private comparison protocol\cite{35}, quantum key agreement(QKA) protocols\cite{36,37}.
	\newline
	Firstly, Hsu\cite{30} used the idea of QSA to developed a quantum secret sharing scheme. In this protocol, the Boss encrypts the secret classical bits with the help of Grover's two-particle quantum state and then sends these two qubits to each of the two agents. The Boss's secret can only be recovered when both the agents combine their qubits and perform some collective unitary operations, where both the agents alone can not reveal Boss's secret. His protocol divides the encoded states into two parts: message states and cheat-detecting states. The Boss randomly prepares the two-particle states as the message state or cheat detecting state and tries to find any possible spying using the cheat-detecting state. Later, Hao et al.\cite{31} discuss the dense coding attack on Hsu's scheme\cite{30} and improved the cheat-detecting mode of his protocol by involving three bases for measuring the qubits by the agents. Experimental QSS protocol based on the four-state Grover's algorithm reported by Hao et al.\cite{38}. Later, the work by Tseng et al.\cite{32} used the similar concept of Grover's algorithm. The two main features of their protocol make it distinct from the above work. The first is no need of quantum memory. The second is that agents reconstruct the secret without performing any unitary operation or quantum measurement.
	\newline
	Although several quantum cryptography protocols are introduced based on the two-qubit Grover's search algorithm, this paper studies the first three-qubit Grover's search algorithm to derive the four-party QSS protocol and provide higher encoding capacity. Furthermore, Grover's search algorithm with three-qubits has already been experimentally demonstrating in nuclear magnetic resonance (NMR) systems\cite{39}. We generalize the Hsu\cite{30} scheme into a one-to-three party QSS scheme by introducing Grover's three-particle quantum state. In contrast to the exiting QSS schemes based on the idea of Grover's search algorithm, the proposed protocol is highly efficient and has a higher capacity to encode the secret. Furthermore, we analyze the practical feasibility of the proposed scheme on the quantum simulator. We use Grover's search algorithm\cite{29} to share the secret and detect illusion by any malicious party. The protocol consists $4$ party: one dealer $D$ and three participants $P_{1},P_{2},P_{3}$. In our protocol, the dealer $D$ transmits an encoded state among three participants  $P_{1},P_{2},P_{3}$ by utilizing Grover's three-particle quantum state. The cooperation of all the participants can only reconstruct the dealer's secret; that is, participants alone can not reveal the dealer's secret. At the time of secret sharing, $D$ remembers two things: first, one of the three participants and at least one may be honest. Second, the honest participant will avert the dishonest one from disrupting the action when they work together.
	\newline
	The rest of the paper is organized as follows. In sect. \ref{PP}, Grover's search algorithm is described and introduced the proposed QSS protocol based on three-qubits Grover's algorithm. In sect. \ref{RD}, simulation analysis is done on the quantum simulator offered by `IBM-quantum experience' and we analyze the security of the scheme by considering four types of attacks. In sect. \ref{CA}, the proposed protocol is compared with other exiting schemes. Finally, the conclusion is given in sect. \ref{CO}.

	\section{Proposed QSS protocol}
	\label{PP}
	\subsection{Protocol Design}
	\label{PD}
	Grover's algorithm\cite{29} was proposed for searching an unstructured database with $2^{n}$ elements, in which the database is signified  as an $n$-particle quantum state $|S\rangle= [(1/\sqrt{2})(|0 \rangle +|1\rangle)]^{\otimes n}$ and the searched target $|m\rangle$ is one explicit measurement result in $|S\rangle$. Grover's algorithm consists of the following steps:
	\begin{enumerate}
		\item Consider an initial state $|S\rangle$.
		\item Apply two unitary transformations $U_{m} = I-2|m\rangle \langle m|$ and $U_{S} = 2|S\rangle \langle S|-I$ on the initial state $|S\rangle$. These two operations $U_{m}$ and $U_{S}$ must be executed  $r= \text{round}(\frac{\pi}{4}\sqrt{2^n})$ times.
		\item After the execution of two unitary operations $U_{m}$ and $U_{S}$ on the state $|S\rangle$, the measurement result becomes $|m\rangle$ with a probability close to $1$.
		
	\end{enumerate} 
	Furthermore, after repetitively performing the operations $U_{m}$ and $U_{S}$ on $|S\rangle$, the state $|S\rangle$ approach to the marked state (search target) $|m\rangle$. For instance, let $|S\rangle =|+\rangle|+\rangle|+\rangle = [(1/\sqrt{2})(|0 \rangle +|1\rangle)]^{\otimes 3}$ and the search target (marked state) $|m\rangle$ can be any element from the set \{$|000\rangle,|001\rangle,|010\rangle,|100\rangle,|011\rangle,|101\rangle,|110\rangle,|111\rangle$\}. Then, after the execution of Grover's algorithm, the measurement result of $|S\rangle$ will be the marked state $|m\rangle$ with a probability close to $1$.
	\newline	
	We present a $(3,3)$ threshold quantum secret sharing scheme based on a three-qubit Grover's algorithm. The protocol consists of one dealer $D$ and three participants $P_{1},P_{2},P_{3}$. The protocol is to share the secret $s$ as a binary string of length $9$ by repeating the protocol three times. For instance, let $D$ wants to share the secret $s=413=110011101$, then $D$ sends the secret $s$ by generating the shares $s_{1},s_{2},s_{3}$, each share is a binary string of length $3$.
	\\ 
	The proposed protocol is accomplished with an initial state $|S_{k}\rangle$ as the input to the Grover's algorithm. The dealer $D$ prepares this initial state $|S_{k}\rangle$ by taking each particle in $|S_{k}\rangle$ from any one of the following four eigenstates:
	\begin{equation*}
		|+\rangle = (1/\sqrt{2})(|0 \rangle +|1\rangle), \;|-\rangle = (1/\sqrt{2})(|0 \rangle -|1\rangle),$$ $$ |+i\rangle = (1/\sqrt{2})(|0 \rangle +i|1\rangle), \;|-i\rangle = (1/\sqrt{2})(|0 \rangle -i|1\rangle).
	\end{equation*}	
	In this study, the encoded quantum state is divided into two categories: message transmitting states and  cheat-detecting states. The dealer $D$ encodes the shares $s_{1},s_{2}\; \text{and} \;s_{3}$ as the marked states $|110\rangle, |011\rangle$ and $|101\rangle$, respectively. $D$ encrypts his secret shares in the state $U_{m}|S_{k}\rangle$ by performing the unitary operation $U_{m}$ on the initial state $|S_{k}\rangle$, where $m$ can be either  $110,~011$ or $101$. Detection of a possible eavesdropping is done by harnessing the state $U_{m}|S_{k}\rangle$, where $m$ can be an element from the set \{$000,001,010,100,111$\}. The process of the proposed secret-sharing scheme based on the three-qubits Grover's algorithm involves the following six steps:
	\begin{enumerate}
		\item The dealer $D$ randomly prepares a three-particle initial state $|S_{k}\rangle \in \Omega,$ for $k=1,2,...,64$. $D$ performs the operation $U_{m}$ on $|S_{k}\rangle$ so as to obtain an encoded state. All encoded states $ U_{m}|S_{k}\rangle$ ( say $|S_{k}\rangle_{m}$) are maximally entangled states and $\Omega$ is the set of all the $64$ initial states listed in Table \ref{tab.1}.
		
		\item The dealer $D$ transmits one of the three qubits to each of the participants $P_{1}, P_{2}$ and $P_{3}$. It is assumed that $P_{1}$ receives the first, $P_{2}$ receives the second, and $P_{3}$ receives the third qubit, respectively.

		\item Also, before transmitting $D$ confirms that every participant receives his qubit via classical channel.

		\item  After the verification that all the participants received their respective qubit, $D$ publicly declares his initial state formulation $|S_{k}\rangle$.
		
		\item The participants $P_{1},P_{2}$ and $P_{3}$ amalgamate their qubits and decode the secret by executing the collective unitary operations $U_{S_{k}},U_{M}$ and $U_{S_{k}}$ respectively. Firstly, they operate the transformation $U_{S_{k}}$ on these three-qubits and observe the measurement outcome of the state, which is the marked state say $|M\rangle$ with higher probability. Now, they operate the two transformations $U_{M}$ and $U_{S_{k}}$ on these three-qubits, respectively. They determine the measurement result of these three-qubits will be the marked state $|m\rangle$ with probability close to $1$. For simplicity the operations $U_{S_{k}},U_{M},$ and $U_{S_{k}}$ are denoted by single operation $U_{S_{k}MS_{k}}$.	
		
		\begin{figure*}[htbp!]\centering
			\includegraphics[width=9cm,height=6.5cm]{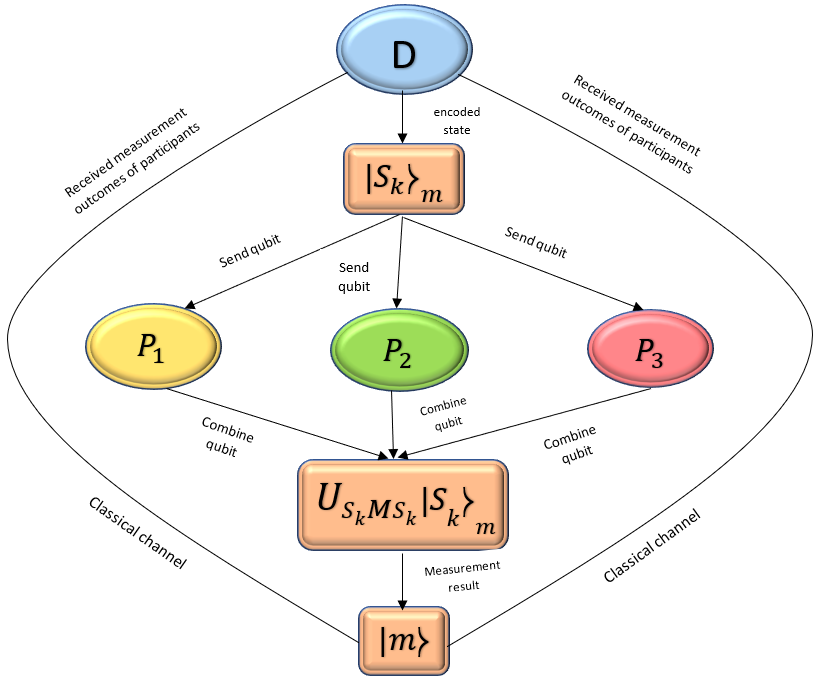}
			\caption{Proposed scheme for QSS}	
			\label{fig1}
		\end{figure*}
		
		\item The participants $P_{1},P_{2}$ and $P_{3}$ get their outcomes by performing the respective local measurements in the computational bases. The protocol terminates successfully when each participant informs the dealer $D$ about their measurement result over a secure classical channel. The schematic of the proposed scheme is given in Fig. \ref{fig1}.
		
	\end{enumerate}

	\subsection{Protocol implementation}
	\label{PI}
	The proposed protocol is implemented in the following steps:
	\begin{enumerate}
		\item Suppose, the dealer $D$ encodes the three-bits information $110$ as the marked state $|m\rangle = |110\rangle$ and prepares the initial state $|S_{1}\rangle = |+\rangle |+\rangle |+\rangle  = [(1/\sqrt{2})(|0 \rangle +|1\rangle)]^{\otimes3}$. Then, $D$ performs the operation $U_{m}$ on $|S_{1}\rangle$ so as to obtain the encoded state $|S_{1}\rangle_{m}$.
		
		\item  $D$ transmits one of the three-qubits to each participants $P_{1},P_{2}$ and $P_{3}$.

		\item  $D$ does not broadcast the initial formulation until it has verified that all three participants have received their respective qubits. $D$ expects the honest participant to announce a message publicly. Therefore, $D$ knows whether the honest participant has received a qubit; the dishonest participant(s) also announce that they have received a qubit, failing which $D$ does not disclose his formulation.

		\item  $D$ publicly announces his initial state formulation as he verified that the honest participant had received a qubit.

		\item After receiving the information of initial state formulation, the honest participant(s) can combine his(their) qubit with the other qubits to decode the secret share by performing the unitary transformation $U_{S_{k}MS_{k}}$.  
		
		\item The participants $P_{1},P_{2}$ and $P_{3}$ combine their qubits, and by operating $U_{S_{1}}$ on these three-qubits get the resulting state as follow:
		\begin{align}\nonumber
			U_{S_{1}}|S_{1}\rangle_{110} & = \dfrac{1}{2\sqrt{8}} \Big[|000\rangle +|001\rangle+|010\rangle+|011\rangle+|100\rangle+|101\rangle+ 5|110\rangle \\ & \quad\quad +|111\rangle\Big].
		\end{align}
		It is known that the probability of the outcome state $|110\rangle$ is higher than the other states, so all the participants perform the two transformations $U_{M=110}$ and $U_{S_{1}}$, thereby resulting in the state as follows:
		\begin{align}\nonumber
			U_{S_{1}}U_{110}U_{S_{1}}|S_{1}\rangle_{110} & = \dfrac{1}{4\sqrt{8}} \Big[-|000\rangle -|001\rangle-|010\rangle-|011\rangle-|100\rangle-|101\rangle
			\\ & \quad\quad  +11 |110\rangle-|111\rangle\Big].
		\end{align}
		Hence from the equation $(2)$, $P_{1},P_{2}$ and $P_{3}$  get the measurement result of the encoded state as the marked state $|110\rangle$ with probability $0.9453$.

		\item  $P_{1},P_{2}$ and $P_{3}$ execute the collective operation $U_{S_{k}MS_{k}}$ together, thereby preventing the dishonest participant(s) from damaging the share. In other words, the dishonest participant(s) cannot perform any deception in the collective operation $U_{S_{k}MS_{k}}$.

		\item After the operation $U_{S_{k}MS_{k}}$ is performed, all the three participants $P_{1},P_{2}$ and $P_{3}$ have to perform the respective local measurements in the computational basis. In this case $P_{1},P_{2}$ and $P_{3}$ inform $D$ about the secret shares $s_{1}$ if their outcomes are $1,1$ and $0$, respectively. No further discussion is needed after they have made their own local measurements. However, one or two participant(s) may be dishonest. Therefore, $D$ and the trustworthy participant(s) should devise a mechanism to check any malicious activity.

		\item  $D$ at random determines whether to use the three-qubit state as the message state or the cheat-detecting state. $D$ encodes his three-bit information into the three-qubit state if the marked state $|m\rangle$ is either $|110\rangle,|011\rangle$ or $|101\rangle$, and when the marked state is any one of the states $|000\rangle, |001\rangle,|010\rangle,|100\rangle$ or $|111\rangle$, $D$ does not encrypt any secret information. By using cheat-detecting codes, the dealer $D$ can detect the possible eavesdropping. 
		
		\item $D$ prepares the cheat-detecting states, and he expects that all the participants publicize their outcomes. If $D$ found that there is no eavesdropping on the resulting state or outcomes which are publicly announced by $P_{1},P_{2}$ and $P_{3}$, then $D$ decides to send the shares of the secret $s$. Otherwise, he can publicly tell $P_{1},P_{2}$ and $P_{3}$ what has been deceived. Therefore, $D$ will be aware of any deception that disturbs the result.
		
	\end{enumerate}

	\section{Results and Discussions}
	\label{RD}
	\subsection{Simulation analysis}
	\label{SA}
\begin{figure*}[htbp!]
		\includegraphics[width=12cm,height=3.5cm]{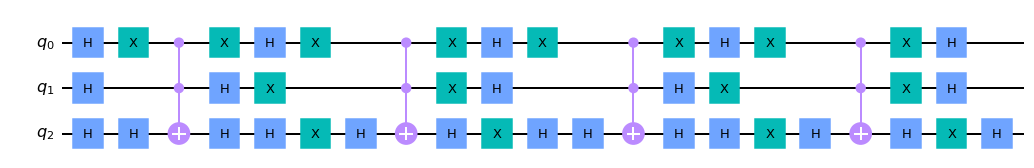}\\
		\caption{Quantum circuit of proposed scheme with $|S_{1}\rangle=|+++\rangle$ and $|m\rangle=|110\rangle$} 
		\label{fig2}
	\end{figure*}

\begin{figure*}[htbp!]
		\centering
		\begin{minipage}[b]{0.45\linewidth}
			\includegraphics[width=6cm,height=4cm]{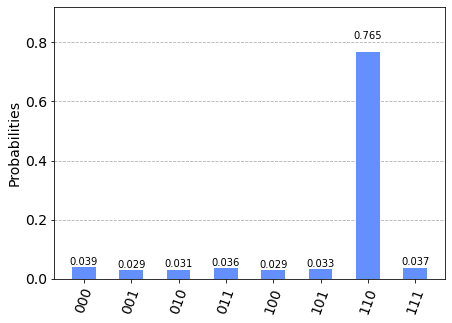}
			\caption{Probability of the outcomes of $U_{S_{1}}|S_{1} \rangle_{110.}$}
			\label{fig3}
		\end{minipage}
		\quad
		\begin{minipage}[b]{0.45\linewidth}
			\includegraphics[width=6cm,height=4cm]{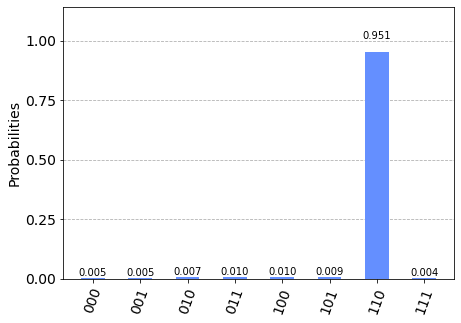}
			\caption{Probability of the outcomes of $U_{S_{1}MS_{1}}|S_{1} \rangle_{110.}$}
			\label{fig4}
		\end{minipage}
	\end{figure*}

	Based on Grover's algorithm, we test the proposed QSS scheme on the quantum simulator offered by IBM on its cloud server `IBM-Quantum Experience'\cite{40} to demonstrate the scheme's experimental viability. The quantum simulator is $'ibmq\_qasm\_simulator'$.  The scheme is based on the measurement result of the outcome state, which is demonstrated by the probability of the marked state $|110\rangle$. The quantum circuit on the $'qasm\_simulator'$ of the proposed scheme with $|S_{1}\rangle = |+\rangle |+\rangle |+\rangle$ initial state and  $|110\rangle$ marked state is shown in Fig. \ref{fig2}. The probabilities of the states after the execution of the operators  $U_{S_{1}}|S_{1}\rangle_{110}$ and 	$U_{S_{1}}U_{M}U_{S_{1}}|S_{1}\rangle_{110}$ given in Fig. \ref{fig3} and Fig. \ref{fig4}, respectively.
	\newline
	
	
	As it can be seen from Fig. \ref{fig3} and Fig. \ref{fig4}, the measurement outcome of the state $U_{S_{1}}|S_{1} \rangle_{110}$ is $|110\rangle$ with probability $0.765$ and by observing the marked state $|M\rangle=|110\rangle$ the measurement outcome of the state $U_{S_{1}MS_{1}}|S_{1} \rangle_{110}$ is marked state $|m\rangle=|110\rangle$ with probability $0.951$.		
	
	\subsection{Security analysis}
	\label{SEA}
	The security analysis of the proposed scheme is based on the dealer's initial preparation and the participant's outcomes. $D$ tries to find every possible eavesdropping of the malicious participants by using cheat-detecting codes and measuring the resulting state\cite{30}. Here, we will consider four types of attack: participants lie, intercept attack, intercept and resend attack, and entangled-measure attack. In the following analysis, we showed that the proposed scheme is secure against all these attacks.

	\subsubsection{Participants lie about their measurement outcomes}
	\label{PL}
	Suppose $P_{1}$ and $P_{2}$ are dishonest participants. In step 6 of sect.\ref{PD}, all the participants $P_{1},P_{2}$ and $P_{3}$ have to inform $D$ about their own measurement outcomes. $P_{1}$ and $P_{2}$ could have intentionally misinformed $D$ about their outcomes. However, $D$ detects this type of deception since it alters the encoded state's measurement result. For illustration, let the marked state is $|m\rangle=|101\rangle$. So, after performing their own local measurements participants $P_{1},P_{2}$ and $P_{3}$ get the outcomes $1,0$ and $1$, respectively. If dishonest participants $P_{1}$ and $P_{2}$ inform $D$ that their respective outcomes are $0$ and $1$ than the marked state $|m\rangle$ become $|011\rangle$. Which is different from $D$'s marked state. Therefore, the dishonest participant(s) can not perform any deception by declaring fake outcomes or by making no declaration.

	\subsubsection{Intercept attack}
	\label{IA}
	In this attack, dishonest participant(s) want to eavesdrop on the qubits of the other participants. We will discuss intercept-attack by including two cases of initial state preparation; first for instance, $D$ prepares either $|S_{1}\rangle_{m}$ or $|S_{9}\rangle_{m}$ and second, $D$ can prepares any possible state $|S_{k}\rangle_{m}$, where $k=1,2,...,64.$
	\newline
	For instance, $D$ prepared either $|S_{1}\rangle_{110}$ or $|S_{9}\rangle_{110}$ initial state. Let $P_{1}$ is a dishonest participant, which purpose is to identify $D$'s secret without $P_{2}$ and $P_{3}$ cooperation. The dishonest participant $P_{1}$ intercepts $P_{2}$'s and $P_{3}$'s qubits and then operate either $U_{S_{1}MS_{1}}$ or $U_{S_{9}MS_{9}}$ on the three qubits. If $P_{1}$ chooses the wrong operation $U_{S_{k}MS_{K}}$ then the resulting states are shown as follows: 
	\begin{align}\nonumber
		U_{S_{9}}|S_{1} \rangle_{110}= Q &= \dfrac{-1}{2\sqrt{8}}\Big[(2+i)|000\rangle+|001\rangle+|010\rangle+(2-i)|011\rangle+|100\rangle \\ & \quad\quad\quad\quad\quad +(2-i)|101\rangle  -(2+i)|110\rangle+3|111\rangle\Big],
	\end{align}	
	\begin{align}\nonumber
		U_{S_{1}}|S_{9} \rangle_{110}= R &= \dfrac{-1}{2\sqrt{8}}\Big[(-2+i)|000\rangle-i|001\rangle-i|010\rangle+(2+i)|011\rangle-i|100\rangle \\ & \quad\quad\quad\quad\quad +(2+i)|101\rangle+(-2+i)|110\rangle+3i|111\rangle\Big],	
	\end{align} 
	from $(3)$ and $(4)$, we observe that the probability of the measurement outcome  $|111\rangle$ is higher than the other states. If $P_{1}$ take the marked state $|M\rangle = |111\rangle$ and operate the transformations $U_{M}$ and $U_{S_{k}}$ sequentially on states $Q$ and $R$, then the states become
	\begin{align}\nonumber
		U_{S_{9}}U_{111}Q &= \dfrac{1}{4\sqrt{8}}\Big[(4+3i)|000\rangle+|001\rangle+|010\rangle+(4-3i)|011\rangle+|100\rangle \\ & \quad\quad\quad\quad\quad +(4-3i)|101\rangle-(4+3i)|110\rangle-5|111\rangle\Big],
	\end{align}   
	\begin{align}\nonumber
		U_{S_{1}}U_{111}R &= \dfrac{1}{4\sqrt{8}}\Big[(-4+3i)|000\rangle-i|001\rangle-i|010\rangle+(4+3i)|011\rangle-i|100\rangle \\ & \quad\quad\quad\quad\quad +(4-3i)|101\rangle+(-4+3i)|110\rangle-5i|111\rangle\Big].
	\end{align}
	From $(5)$ and $(6)$, we see that the measurement outcomes of the states  are $|000\rangle,|011\rangle,$ $|101\rangle,|110\rangle$ and $|111\rangle$ with probability $0.195$. If $P_{1}$ assume the correct marked state $|M\rangle = |110\rangle$ but performs wrong operations $U_{S_{k}}$ on states $Q$ and $R$ then the resulting states are
	\begin{align}\nonumber
		U_{S_{9}}U_{110}Q &= \dfrac{1}{4\sqrt{8}}\Big[(6+i)|000\rangle+(3+2i)|001\rangle+(3+2i)|010\rangle+(2-i)|011\rangle
		\\ & \quad\quad\quad +(3+2i)|100\rangle+(2-i)|101\rangle+(2+3i)|110\rangle+(5-2i)|111\rangle\Big],
	\end{align}   
	\begin{align}\nonumber
		U_{S_{1}}U_{110}R &= \dfrac{1}{4\sqrt{8}}\Big[(6-i)|000\rangle+(2+3i)|001\rangle+(2+3i)|010\rangle-(2+i)|011\rangle
		\\ & \quad\quad\quad +(2+3i) |100\rangle-(2+i)|101\rangle+(-2+3i)|110\rangle+(2-5i)|111\rangle\Big].
	\end{align}
	The measurement outcomes of the states $(7)$ and $(8)$ are $|000\rangle$ with probability $0.289$. Hence, the above quantum states show that the dishonest participant $P_{1}$ can not gain any secret information to operate the wrong collective operation $U_{S_{k}MS_{k}}$. Assuming that $D$ prepared cheat-detecting states or message states with equal probability, and the dishonest participant operates $U_{S_{1}MS_{1}}$ or $U_{S_{9}MS_{9}}$ with equal probability, the dishonest participant $P_{1}$ assumes the correct marked state $|M\rangle$ with probability $\frac{1}{8}$. Then, $P_{1}$ can access the secret with probability $\frac{1}{32}$. Moreover, some conditions in which  $D$ can instantly detect deception:(i) $D$ finds the resulting states different from his prepared states. (ii) $D$ prepares cheat-detecting state and honest participant find cheat-detecting state. (iii) $D$ and honest participant find the approximate probability of the marked state $|m\rangle$ less than $1$.

	\begin{table}[htbp]%
		
		\caption{The $64$ measurement outcomes (Motcm) with maximum probability $(P)$ of the states $U_{S_{k}}|S_{1}\rangle_{110},$ where $k=1,2,...,64.$ After observing the measurement result i.e. marked state $|M\rangle$ of the state $U_{S_{k}}|S_{1}\rangle_{110}$, the measurement outcomes (Motcm) with maximum probability $(P)$ of the states $U_{S_{k}}U_{M}U_{S_{k}}|S_{1}\rangle_{110},$ where $k=1,2,...,64$. The measurement outcomes must be the marked state $|M\rangle=|110\rangle$, and $|m\rangle=|110\rangle$  with probability $0.781$ and $0.945$, respectively only when $k=1$.}
		\label{tab.1}
		\label{aggiungi}\centering\small
		\renewcommand{\arraystretch}{-0.5}
		\resizebox{11cm}{!}
		{\begin{tabular}{lllllll}
				\midrule
				
				$k$ &  $|S_{k}\rangle$ & $\text{Motcm}(U_{S_{k}}|S_{1}\rangle_{110})$  & $P$ & $|M\rangle$ & $\text{Motcm}(U_{S_{k}}U_{M}U_{S_{k}}|S_{1}\rangle_{110})$ & $P$ \\ \toprule
				
				$1$ & $|+\rangle|+\rangle|+\rangle$ & $|110\rangle$ & $0.781$ & $|110\rangle$ & $|110\rangle$ & $0.945$ \\

				$2$ & $|+\rangle|+\rangle|-\rangle$ & $|000\rangle,|010\rangle,|100\rangle$ & $0.281$ & $|000\rangle$ & $|010\rangle,|100\rangle$ & $0.383$\\

				$3$ & $|-\rangle|+\rangle|+\rangle$ & $|100\rangle,|101\rangle,|111\rangle$ & $0.281$ & $|100\rangle$ & $|101\rangle,|111\rangle$ & $0.383$ \\
				
				$4$ & $|+\rangle|-\rangle|+\rangle$ & $|010\rangle,|011\rangle,|111\rangle$ & $0.281$ & $|010\rangle$ & $|011\rangle,|111\rangle$ & $0.383$\\

				$5$ & $|+\rangle|-\rangle|-\rangle$ & $|001\rangle,|010\rangle,|101\rangle$ & $0.281$ & $|001\rangle$ & $|010\rangle,|101\rangle$ & $0.383$\\

				$6$ & $|-\rangle|+\rangle|-\rangle$ & $|001\rangle,|011\rangle,|100\rangle$ & $0.281$ & $|001\rangle$ & $|011\rangle,|100\rangle$ & $0.383$\\

				$7$ & $|-\rangle|-\rangle|+\rangle$ & $|000\rangle,|001\rangle,|111\rangle$ & $0.281$ & $|001\rangle$ & $|000\rangle,|111\rangle$ & $0.383$\\

				$8$ & $|-\rangle|-\rangle|-\rangle$ & $|000\rangle,|011\rangle,|101\rangle$ & $0.281$ & $|011\rangle$ & $|000\rangle,|101\rangle$ & $0.383$\\

				$9$ & $|+i\rangle|+i\rangle|+i\rangle$  & $|111\rangle$ & $0.281$ & $|111\rangle$ & $|000\rangle,|011\rangle,|101\rangle,|110\rangle,|111\rangle$ & $0.195$\\

				$10$ & $|+i\rangle|+i\rangle|-i\rangle$  & $|001\rangle$ & $0.406$ & $|001\rangle$ & $|001\rangle$ & $0.477$ \\

				$11$ & $|-i\rangle|+i\rangle|+i\rangle$  & $|100\rangle$ & $0.281$ & $|100\rangle$ & $|000\rangle,|011\rangle,|100\rangle,|101\rangle,|110\rangle$ & $0.195$\\

				$12$ & $|+i\rangle|-i\rangle|+i\rangle$  & $|010\rangle$ & $0.281$ & $|010\rangle$ & $|000\rangle,|010\rangle,|011\rangle,|101\rangle,|110\rangle$ & $0.195$ \\

				$13$ & $|+i\rangle|-i\rangle|-i\rangle$  & $|100\rangle$ & $0.281$ & $|100\rangle$ & $|000\rangle,|011\rangle,|100\rangle,|101\rangle,|110\rangle$ & $0.195$\\

				$14$ & $|-i\rangle|+i\rangle|-i\rangle$  & $|010\rangle$ & $0.281$ & $|010\rangle$ & $|000\rangle,|010\rangle,|011\rangle,|101\rangle,|110\rangle$ & $0.195$\\

				$15$ & $|-i\rangle|-i\rangle|+i\rangle$  & $|001\rangle$ & $0.406$ & $|001\rangle$ & $|001\rangle$ & $0.477$\\

				$16$ & $|-i\rangle|-i\rangle|-i\rangle$  & $|111\rangle$ & $0.281$  & $|111\rangle$ & $|000\rangle,|011\rangle,|101\rangle,|110\rangle,|111\rangle$ & $0.195$\\

				$17$ & $|+\rangle|+\rangle|+i\rangle$  & $|110\rangle$ & $0.406$ & $|110\rangle$ & $|110\rangle$ & $0.477$\\

				$18$ & $|+\rangle|+\rangle|-i\rangle$  & $|110\rangle$ & $0.406$ & $|110\rangle$ & $|110\rangle$ & $0.477$\\

				$19$ & $|-\rangle|-\rangle|+i\rangle$  & $|000\rangle$ & $0.281$ & $|000\rangle$ & $|000\rangle,|001\rangle,|011\rangle,|101\rangle,|111\rangle$ & $0.195$\\

				$20$ & $|-\rangle|-\rangle|-i\rangle$  & $|000\rangle$ & $0.281$ & $|000\rangle$ & $|000\rangle,|001\rangle,|011\rangle,|101\rangle,|111\rangle$ & $0.195$\\

				$21$ & $|+\rangle|-\rangle|+i\rangle$  & $|010\rangle$ & $0.281$ & $|010\rangle$ & $|001\rangle,|010\rangle,|011\rangle,|101\rangle,|111\rangle$ & $0.195$\\

				$22$ & $|+\rangle|-\rangle|-i\rangle$  & $|010\rangle$ & $0.281$ & $|010\rangle$ & $|001\rangle,|010\rangle,|011\rangle,|101\rangle,|111\rangle$ & $0.195$\\

				$23$ & $|+\rangle|+i\rangle|-\rangle$  & $|010\rangle$ & $0.281$ & $|010\rangle$ & $|000\rangle,|001\rangle,|010\rangle,|100\rangle,|101\rangle$ & $0.195$\\

				$24$ & $|+\rangle|-i\rangle|-\rangle$  & $|010\rangle$ & $0.281$ & $|010\rangle$ & $|000\rangle,|001\rangle,|010\rangle,|100\rangle,|101\rangle$ & $0.195$\\

				$25$ & $|-\rangle|+\rangle|+i\rangle$  & $|100\rangle$ & $0.281$ & $|100\rangle$ & $|001\rangle,|011\rangle,|100\rangle,|101\rangle,|111\rangle$ & $0.195$\\

				$26$ & $|-\rangle|+\rangle|-i\rangle$  & $|100\rangle$ & $0.281$ & $|100\rangle$ & $|001\rangle,|011\rangle,|100\rangle,|101\rangle,|111\rangle$ & $0.195$\\

				$27$ & $|-\rangle|+i\rangle|+\rangle$  & $|111\rangle$ & $0.281$ & $|111\rangle$ & $|000\rangle,|001\rangle,|100\rangle,|101\rangle,|111\rangle$ & $0.195$\\

				$28$ & $|-\rangle|-i\rangle|+\rangle$  & $|111\rangle$ & $0.281$ & $|111\rangle$ & $|000\rangle,|001\rangle,|100\rangle,|101\rangle,|111\rangle$ & $0.195$\\

				$29$ & $|+i\rangle|+\rangle|+i\rangle$  & $|110\rangle$ & $0.281$ & $|110\rangle$ & $|000\rangle,|010\rangle,|101\rangle,|110\rangle,|111\rangle$ & $0.195$\\

				$30$ & $|+i\rangle|-\rangle|+\rangle$  & $|111\rangle$ & $0.281$ & $|111\rangle$ & $|000\rangle,|001\rangle,|010\rangle,|011\rangle,|111\rangle$ & $0.195$\\

				$31$ & $|-i\rangle|+\rangle|-\rangle$  & $|100\rangle$ & $0.281$ & $|100\rangle$ & $|000\rangle,|001\rangle,|010\rangle,|011\rangle,|100\rangle$ & $0.195$\\

				$32$ & $|-i\rangle|-\rangle|+\rangle$  & $|111\rangle$ & $0.281$ & $|111\rangle$ & $|000\rangle,|001\rangle,|010\rangle,|011\rangle,|111\rangle$ & $0.195$\\

				$33$ & $|+i\rangle|+i\rangle|+\rangle$ & $|111\rangle$ & $0.406$ & $|111\rangle$ & $|111\rangle$ & $0.477$\\

				$34$ & $|+i\rangle|+i\rangle|-\rangle$ & $|001\rangle$ & $0.281$ & $|001\rangle$ & $|001\rangle,|010\rangle,|011\rangle,|100\rangle,|101\rangle$ & $0.195$\\
				
				$35$ & $|-i\rangle|-i\rangle|+\rangle$ & $|111\rangle$ & $0.406$ & $|111\rangle$ & $|111\rangle$ & $0.477$\\
				
				$36$ & $|-i\rangle|-i\rangle|-\rangle$ & $|001\rangle$ & $0.281$ & $|001\rangle$ & $|001\rangle,|010\rangle,|011\rangle,|100\rangle,|101\rangle$ & $0.195$\\
				
				$37$ & $|+\rangle|+i\rangle|-i\rangle$ & $|010\rangle$ & $0.406$ & $|010\rangle$ & $|010\rangle$ & $0.477$\\
				
				$38$ & $|+\rangle|-i\rangle|+i\rangle$ & $|010\rangle$ & $0.406$ & $|010\rangle$ & $|010\rangle$ & $0.477$\\
				
				$39$ & $|-\rangle|+i\rangle|-i\rangle$ & $|001\rangle$ & $0.281$ & $|001\rangle$ & $|000\rangle,|001\rangle,|011\rangle,|100\rangle,|111\rangle$ & $0.195$\\
				
				$40$ & $|-\rangle|-i\rangle|+i\rangle$ & $|001\rangle$ & $0.281$ & $|001\rangle$ & $|000\rangle,|001\rangle,|011\rangle,|100\rangle,|111\rangle$ & $0.195$\\
				
				$41$ & $|+i\rangle|+\rangle|-i\rangle$ & $|100\rangle$ & $0.406$ & $|100\rangle$ & $|100\rangle$ & $0.477$\\
				
				$42$ & $|+i\rangle|-\rangle|-i\rangle$ & $|001\rangle$ & $0.281$ & $|001\rangle$ & $|000\rangle,|001\rangle,|010\rangle,|101\rangle,|111\rangle$ & $0.195$\\
				
				$43$ & $|+i\rangle|-i\rangle|+\rangle$ & $|110\rangle$ & $0.281$ & $|110\rangle$ & $|010\rangle,|011\rangle,|100\rangle,|101\rangle,|110\rangle$ & $0.195$\\
				
				$44$ & $|+i\rangle|-i\rangle|-\rangle$ & $|000\rangle$ & $0.281$ & $|000\rangle$ & $|000\rangle,|010\rangle,|011\rangle,|100\rangle,|101\rangle$ & $0.195$\\
				
				$45$ & $|-i\rangle|+\rangle|+i\rangle$ & $|100\rangle$ & $0.406$ & $|100\rangle$ & $|100\rangle$ & $0.477$\\
				
				$46$ & $|-i\rangle|-\rangle|+i\rangle$ & $|001\rangle$ & $0.281$ & $|001\rangle$ & $|000\rangle,|001\rangle,|010\rangle,|101\rangle,|111\rangle$ & $0.195$\\
				
				$47$ & $|-i\rangle|+i\rangle|+\rangle$ & $|110\rangle$ & $0.281$ & $|110\rangle$ & $|010\rangle,|011\rangle,|100\rangle,|101\rangle,|110\rangle$ & $0.195$\\
				
				$48$ & $|-i\rangle|+i\rangle|-\rangle$ & $|000\rangle$ & $0.281$ & $|000\rangle$ & $|000\rangle,|010\rangle,|011\rangle,|100\rangle,|101\rangle$ & $0.195$\\
				
				$49$ & $|+\rangle|+i\rangle|+i\rangle$ & $|110\rangle$ & $0.281$ & $|110\rangle$ & $|000\rangle,|011\rangle,|100\rangle,|110\rangle,|111\rangle$ & $0.195$\\
				
				$50$ & $|+\rangle|+i\rangle|+\rangle$ & $|110\rangle$ & $0.406$ & $|110\rangle$ & $|110\rangle$ & $0.477$\\
				
				$51$ & $|+\rangle|-i\rangle|-i\rangle$ & $|110\rangle$ & $0.281$ & $|110\rangle$ & $|000\rangle,|011\rangle,|100\rangle,|110\rangle,|111\rangle$ & $0.195$\\
				
				$52$ & $|+\rangle|-i\rangle|+\rangle$ & $|110\rangle$ & $0.406$ & $|110\rangle$ & $|110\rangle$ & $0.477$\\
				
				$53$ & $|-\rangle|+i\rangle|+i\rangle$ & $|101\rangle$ & $0.281$ & $|101\rangle$ & $|000\rangle,|011\rangle,|100\rangle,|101\rangle,|111\rangle$ & $0.195$\\
				
				$54$ & $|-\rangle|+i\rangle|-\rangle$ & $|011\rangle$ & $0.281$ & $|011\rangle$ & $|000\rangle,|001\rangle,|011\rangle,|100\rangle,|101\rangle$ & $0.195$\\
				
				$55$ & $|-\rangle|-i\rangle|-i\rangle$ & $|101\rangle$ & $0.281$ & $|101\rangle$ & $|000\rangle,|011\rangle,|100\rangle,|101\rangle,|111\rangle$ & $0.195$\\
				
				$56$ & $|-\rangle|-i\rangle|-\rangle$ & $|011\rangle$ & $0.281$ & $|011\rangle$ & $|000\rangle,|001\rangle,|011\rangle,|100\rangle,|101\rangle$ & $0.195$\\
				
				$57$ & $|+i\rangle|+\rangle|+\rangle$ & $|110\rangle$ & $0.406$ & $|110\rangle$ & $|110\rangle$ & $0.477$\\
				
				$58$ & $|+i\rangle|+\rangle|-\rangle$ & $|100\rangle$ & $0.281$ & $|100\rangle$ & $|000\rangle,|001\rangle,|010\rangle,|011\rangle,|100\rangle$ & $0.195$\\
				
				$59$ & $|+i\rangle|-\rangle|-\rangle$ & $|101\rangle$ & $0.281$ & $|101\rangle$ & $|000\rangle,|001\rangle,|010\rangle,|011\rangle,|101\rangle$ & $0.195$\\
				
				$60$ & $|+i\rangle|-\rangle|+i\rangle$ & $|011\rangle$ & $0.281$ & $|011\rangle$ & $|000\rangle,|010\rangle,|011\rangle,|101\rangle,|111\rangle$ & $0.195$\\
				
				$61$ & $|-i\rangle|+\rangle|+\rangle$ & $|110\rangle$ & $0.406$ & $|110\rangle$ & $|110\rangle$ & $0.477$\\
				
				$62$ & $|-i\rangle|+\rangle|-i\rangle$ & $|110\rangle$ & $0.281$ & $|110\rangle$ & $|000\rangle,|010\rangle,|101\rangle,|110\rangle,|111\rangle$ & $0.195$\\
				
				$63$ & $|-i\rangle|-\rangle|-\rangle$ & $|101\rangle$ & $0.281$ & $|101\rangle$ & $|000\rangle,|001\rangle,|010\rangle,|011\rangle,|101\rangle$ & $0.195$\\
				
				$64$ & $|-i\rangle|-\rangle|-i\rangle$ & $|011\rangle$ & $0.281$ & $|011\rangle$ & $|000\rangle,|010\rangle,|011\rangle,|101\rangle,|111\rangle$ & $0.195$\\\midrule

		\end{tabular}}
	\end{table}

	\begin{table}[htbp]%
		
		\caption{The $64$ measurement outcomes (Motcm) with maximum probability $(P)$ of $U_{S_{k}MS_{k}}|S_{1}\rangle_{110},$ where $k=1,2,...,64$, and $|M\rangle=|110\rangle$. The measurement outcome must be the marked state $|m\rangle=|110\rangle$ with probability $0.951$, only when $k=1$ and $|M\rangle=|110\rangle$.}
		\label{tab.2}

		\label{aggiungi}\centering\small
		\renewcommand{\arraystretch}{-0.5}
		\resizebox{7.5cm}{!}
		{\begin{tabular}{llllllll}
				\midrule
				
				$k$ &  $|S_{k}\rangle$ & $\text{Motcm}(U_{S_{k}MS_{k}}|S_{1}\rangle_{110}) $  & $P$ \\ \toprule
				
				$1$ & $|+\rangle|+\rangle|+\rangle$ & $|110\rangle$ & $0.945$ \\

				$2$ & $|+\rangle|+\rangle|-\rangle$ & $|001\rangle,|011\rangle,|101\rangle,|111\rangle$ & $0.195$  \\

				$3$ & $|-\rangle|+\rangle|+\rangle$ & $|000\rangle,|001\rangle,|010\rangle,|011\rangle$ & $0.195$ \\
				
				$4$ & $|+\rangle|-\rangle|+\rangle$ & $|000\rangle,|001\rangle,|100\rangle,|101\rangle$ & $0.195$ \\

				$5$ & $|+\rangle|-\rangle|-\rangle$ & $|000\rangle,|011\rangle,|100\rangle,|111\rangle$ & $0.195$ \\

				$6$ & $|-\rangle|+\rangle|-\rangle$ & $|000\rangle,|010\rangle,|101\rangle,|111\rangle$ & $0.195$ \\

				$7$ & $|-\rangle|-\rangle|+\rangle$ & $|010\rangle,|011\rangle,|100\rangle,|101\rangle$ & $0.195$ \\

				$8$ & $|-\rangle|-\rangle|-\rangle$ & $|001\rangle,|010\rangle,|100\rangle,|111\rangle$ & $0.195$ \\

				$9$ & $|+i\rangle|+i\rangle|+i\rangle$  & $|000\rangle$ & $0.289$ \\

				$10$ & $|+i\rangle|+i\rangle|-i\rangle$  & $|001\rangle,|110\rangle$ & $0.195$ \\

				$11$ & $|-i\rangle|+i\rangle|+i\rangle$  & $|011\rangle$ & $0.289$ \\

				$12$ & $|+i\rangle|-i\rangle|+i\rangle$  & $|101\rangle$ & $0.289$ \\

				$13$ & $|+i\rangle|-i\rangle|-i\rangle$  & $|011\rangle$ & $0.289$ \\

				$14$ & $|-i\rangle|+i\rangle|-i\rangle$  & $|101\rangle$ & $0.289$ \\

				$15$ & $|-i\rangle|-i\rangle|+i\rangle$  & $|001\rangle,|110\rangle$ & $0.195$ \\

				$16$ & $|-i\rangle|-i\rangle|-i\rangle$  & $|000\rangle$ & $0.289$ \\

				$17$ & $|+\rangle|+\rangle|+i\rangle$  & $|110\rangle$ & $0.477$ \\

				$18$ & $|+\rangle|+\rangle|-i\rangle$  & $|110\rangle$ & $0.477$ \\

				$19$ & $|-\rangle|-\rangle|+i\rangle$  & $|010\rangle,|100\rangle$ & $0.195$ \\

				$20$ & $|-\rangle|-\rangle|-i\rangle$  & $|010\rangle,|100\rangle$ & $0.195$ \\

				$21$ & $|+\rangle|-\rangle|+i\rangle$  & $|000\rangle,|100\rangle$ & $0.195$ \\

				$22$ & $|+\rangle|-\rangle|-i\rangle$  & $|000\rangle,|100\rangle$ & $0.195$ \\

				$23$ & $|+\rangle|+i\rangle|-\rangle$  & $|011\rangle,|111\rangle$ & $0.195$ \\

				$24$ & $|+\rangle|-i\rangle|-\rangle$  & $|011\rangle,|111\rangle$ & $0.195$ \\

				$25$ & $|-\rangle|+\rangle|+i\rangle$  & $|000\rangle,|010\rangle$ & $0.195$ \\

				$26$ & $|-\rangle|+\rangle|-i\rangle$  & $|000\rangle,|010\rangle$ & $0.195$ \\

				$27$ & $|-\rangle|+i\rangle|+\rangle$  & $|010\rangle,|011\rangle$ & $0.195$ \\

				$28$ & $|-\rangle|-i\rangle|+\rangle$  & $|010\rangle,|011\rangle$ & $0.195$ \\

				$29$ & $|+i\rangle|+\rangle|+i\rangle$  & $|000\rangle,|010\rangle,|101\rangle,|110\rangle,|111\rangle$ & $0.195$ \\

				$30$ & $|+i\rangle|-\rangle|+\rangle$  & $|100\rangle,|101\rangle$ & $0.195$ \\

				$31$ & $|-i\rangle|+\rangle|-\rangle$  & $|101\rangle,|111\rangle$ & $0.195$ \\

				$32$ & $|-i\rangle|-\rangle|+\rangle$  & $|100\rangle,|101\rangle$ & $0.195$ \\

				$33$ & $|+i\rangle|+i\rangle|+\rangle$ & $|110\rangle$ & $0.289$ \\

				$34$ & $|+i\rangle|+i\rangle|-\rangle$ & $|000\rangle,|111\rangle$ & $0.195$\\
				
				$35$ & $|-i\rangle|-i\rangle|+\rangle$ & $|110\rangle$ & $0.289$\\
				
				$36$ & $|-i\rangle|-i\rangle|-\rangle$ & $|000\rangle,|111\rangle$ & $0.195$\\
				
				$37$ & $|+\rangle|+i\rangle|-i\rangle$ & $|110\rangle$ & $0.289$\\
				
				$38$ & $|+\rangle|-i\rangle|+i\rangle$ & $|110\rangle$ & $0.289$\\
				
				$39$ & $|-\rangle|+i\rangle|-i\rangle$ & $|010\rangle,|101\rangle$ & $0.195$\\
				
				$40$ & $|-\rangle|-i\rangle|+i\rangle$ & $|010\rangle,|101\rangle$ & $0.195$\\
				
				$41$ & $|+i\rangle|+\rangle|-i\rangle$ & $|110\rangle$ & $0.289$\\
				
				$42$ & $|+i\rangle|-\rangle|-i\rangle$ & $|011\rangle,|100\rangle$ & $0.195$\\
				
				$43$ & $|+i\rangle|-i\rangle|+\rangle$ & $|010\rangle,|011\rangle,|100\rangle,|101\rangle,|110\rangle$ & $0.195$\\
				
				$44$ & $|+i\rangle|-i\rangle|-\rangle$ & $|001\rangle,|111\rangle$ & $0.195$\\
				
				$45$ & $|-i\rangle|+\rangle|+i\rangle$ & $|110\rangle$ & $0.289$\\
				
				$46$ & $|-i\rangle|-\rangle|+i\rangle$ & $|011\rangle,|101\rangle$ & $0.195$\\
				
				$47$ & $|-i\rangle|+i\rangle|+\rangle$ & $|010\rangle,|011\rangle,|100\rangle,|101\rangle,|110\rangle$ & $0.195$\\
				
				$48$ & $|-i\rangle|+i\rangle|-\rangle$ & $|001\rangle,|111\rangle$ & $0.195$\\
				
				$49$ & $|+\rangle|+i\rangle|+i\rangle$ & $|000\rangle,|011\rangle,|100\rangle,|110\rangle,|111\rangle$ & $0.195$\\
				
				$50$ & $|+\rangle|+i\rangle|+\rangle$ & $|110\rangle$ & $0.477$\\
				
				$51$ & $|+\rangle|-i\rangle|-i\rangle$ & $|000\rangle,|011\rangle,|100\rangle,|110\rangle,|111\rangle$ & $0.195$\\
				
				$52$ & $|+\rangle|-i\rangle|+\rangle$ & $|110\rangle$ & $0.477$\\
				
				$53$ & $|-\rangle|+i\rangle|+i\rangle$ & $|001\rangle,|010\rangle$ & $0.195$\\
				
				$54$ & $|-\rangle|+i\rangle|-\rangle$ & $|010\rangle,|111\rangle$ & $0.195$\\
				
				$55$ & $|-\rangle|-i\rangle|-i\rangle$ & $|001\rangle,|010\rangle$ & $0.195$\\
				
				$56$ & $|-\rangle|-i\rangle|-\rangle$ & $|010\rangle,|111\rangle$ & $0.195$\\
				
				$57$ & $|+i\rangle|+\rangle|+\rangle$ & $|110\rangle$ & $0.477$\\
				
				$58$ & $|+i\rangle|+\rangle|-\rangle$ & $|101\rangle,|111\rangle$ & $0.195$\\
				
				$59$ & $|+i\rangle|-\rangle|-\rangle$ & $|100\rangle,|111\rangle$ & $0.195$\\
				
				$60$ & $|+i\rangle|-\rangle|+i\rangle$ & $|001\rangle,|100\rangle$ & $0.195$\\
				
				$61$ & $|-i\rangle|+\rangle|+\rangle$ & $|110\rangle$ & $0.477$\\
				
				$62$ & $|-i\rangle|+\rangle|-i\rangle$ & $|000\rangle,|010\rangle,|101\rangle,|110\rangle,|111\rangle$ & $0.195$ \\
				
				$63$ & $|-i\rangle|-\rangle|-\rangle$ & $|100\rangle,|111\rangle$ & $0.195$\\
				
				$64$ & $|-i\rangle|-\rangle|-i\rangle$ & $|001\rangle,|100\rangle$ & $0.195$\\\midrule

	\end{tabular}}\end{table}
	
	Moreover, the honest participant is able to detect dishonest participant's cheating. $D$ prepares some initial state $|S_{k}\rangle \in \Omega$, arbitrarily and then he operate $U_{m}$ on $|S_{k}\rangle \in \Omega$, where $m$ can be $101,011$ or $110$. The dishonest participant $P_{1}$ intercepts other participant's qubits and he prepares some state $|z\rangle=|S_{k'}\rangle_{m'}$. $P_{1}$ sends the corresponding qubits in $|z\rangle$ to the participants $P_{2}$ and $P_{3}$. Instead, $P_{1}$ does not perform any measurement before $D$'s announcement. For illustration, Let $|z\rangle=|S_{1}\rangle_{110} $. Now all the participants $P_{1},P_{2}$ and $P_{3}$ together operate the correct operation $U_{S_{k}MS_{k}}$ on $|z\rangle$. They observe the measurement outcomes with the high probability of the state $U_{S_{k}}|S_{1}\rangle_{110}$ to determine the marked state $|M\rangle$ as listed in Table \ref{tab.1}. From Table \ref{tab.1}, after observing the marked state $|M\rangle$, the participants  $P_{1},P_{2}$ and $P_{3}$ can access the marked state $|m\rangle$ only with the probability $\frac{9}{32}$ (including the case where the probability of the marked state close to $\frac{1}{2}$ or less than $\frac{1}{2}$). According to Table \ref{tab.1}, the participants  $P_{1},P_{2}$ and $P_{3}$ are able to observe the correct marked state $|M\rangle$ with probability $\frac{13}{64}$. If they assume the marked state $|M\rangle =|110\rangle$ and perform $U_{S_{k}MS_{k}}|S_{1}\rangle_{110}$ then the measurement outcomes with high probability are listed in Table \ref{tab.2}. The participants  $P_{1},P_{2}$ and $P_{3}$ assumed the marked state $|M\rangle$ with probability $\frac{1}{8}$. From Table \ref{tab.2}, it can be easy to verified that $P_{1},P_{2}$ and $P_{3}$ are able to measure cheat-detecting states i.e anyone state $|000\rangle,|001\rangle,|010\rangle,|100\rangle,$ or $|111\rangle$ with the approximate probability less than or equal to $\frac{37}{128}$ by performing $U_{S_{k}MS_{k}}|S_{1}\rangle_{110}$.  Hence, $D$ and the honest participant can instantly detect such type of deception with probability $\frac{23}{32}$. Furthermore, the honest participant also detect such cheating when they operate $U_{S_{k}}$ and observed the marked state $|M\rangle$ with probability less than or equal to $\frac{1}{2}$.\\
	Now, let dishonest participant $P_{1}$ intercept $P_{2}$'s and $P_{3}$'s qubits and performs the measurement on the qubits, after operating some collective operation $U_{S_{k}MS_{k}}$.
	From Table \ref{tab.2}, only when the dishonest participant $P_{1}$ operates the correct operation $U_{S_{k}MS_{k}}$ can he access precisely the marked state with probability close to $1$. $P_{1}$ is assumed that performed any $64$-$U_{S_{k}MS_{k}}$ with equal probability. Then it can be verified that $P_{1}$ determines $D$'s secret with probability $\frac{9}{32}$ (including the cases where the probability of marked state $|m\rangle$ close to $\frac{1}{2}$). However, $D$ detect  $P_{1}$'s cheating with probability $\frac{23}{32}$. The dishonest participant cannot gain any information from intercepted state by making a random guess. Therefore, the analysis shows that the proposed scheme is robust and secure.
	
	\subsubsection{ Intercept-Resend attack}
	\label{IR}
	In this attack the dishonest participant prepares his own state and replaces $D$'s qubits by his own qubits and then send the replaced qubits to the other participants. Suppose the dishonest participant $P_{1}$ prepares a triplet state and sends the qubits of the triplet state to the participants $P_{2}$ and $P_{3}$. As $D$ announces his initial state $S_{k}$ in public, $P_{1}$ instantly operates the necessary transformations on his qubit. Here, eight preparations $|S_{k}\rangle_{000},|S_{k}\rangle_{001},|S_{k}\rangle_{010},|S_{k}\rangle_{011},$
	$|S_{k}\rangle_{100},|S_{k}\rangle_{101},|S_{k}\rangle_{110},$ and $|S_{k}\rangle_{111}$ are possible. $P_{1}$ does not have any knowledge about the marked state $|m\rangle$. Therefore, assumed that $P_{1}$ is able to prepare some $|S_{k}\rangle_{m}$ with probability $\frac{1}{8}$. If $D$ prepares a message state but $P_{1}$ prepares a cheat-detecting state then $D$ detects such deception with probability $\frac{5}{8}$. However, if $D$ prepares a cheat-detecting state and $P_{1}$ prepares any other incorrect state then $D$ can detect such deception with probability $\frac{7}{8}$. On average, $D$ can find such type of cheating with probability $\frac{12}{16}$. Hence, the dishonest participant cannot get any information about $D$'s preparation in advance.
	\newline
	Now, some other strategies of intercept-resend attack are considered. The dishonest participant $P_{1}$ can use the intercept-resend attack with orthogonal measurements\cite{41}. $P_{1}$ is assumed to perform the collective measurement on either basis  
	\begin{align}\nonumber
		&	\Big\{|-+++++++\rangle,|+-++++++\rangle,|++-+++++\rangle,|+++-++++\rangle, \\ &
		|++++-+++\rangle,|+++++-++\rangle,|++++++-+\rangle,|+++++++-\rangle \Big\},
	\end{align}
	or
	\begin{align}\nonumber
		& \Big\{|----(+i)(+i)(+i)(+i)\rangle,|++++(+i)(+i)(+i)(+i)\rangle,|+-++(+i)(-i)(+i)(+i)\rangle, \\ & |+-++(+i)(-i)(+i)(+i)\rangle,|++-+(+i)(+i)(+i)(-i)\rangle,|++-+(+i)(+i)(-i)(+i)\rangle, \nonumber \\ &  |--++(+i)(+i)(-i)(-i)\rangle,|--++(-i)(-i)(+i)(+i)\rangle\Big\},
	\end{align}
	where,\\ $|-+++++++\rangle=\dfrac{1}{\sqrt{8}}\Big(-|000\rangle +|001\rangle+|010\rangle+|011\rangle+|100\rangle+|101\rangle+ |110\rangle+|111\rangle\Big)$\\ and \\ $|----(+i)(+i)(+i)(+i)\rangle =\dfrac{1}{\sqrt{8}}\Big(-|000\rangle -|001\rangle-|010\rangle-|011\rangle+i|100\rangle+i|101\rangle+i|110\rangle
	+i|111\rangle\Big)$ and so on.\\
	For instance, suppose either $|S_{1}\rangle_{m}$ or $|S_{9}\rangle_{m}$ is prepared by $D$ and $P_{1}$ decides to perform the orthogonal measurement in the basis $(9)$. If $D$ prepares the state $|S_{1}\rangle_{m}$ then $P_{1}$ can eavesdrop on the marked state without being detected\cite{42}. However, if $D$ prepares $|S_{9}\rangle_{m}$ then the state projects into any orthogonal measurement basis with equal probability $\frac{1}{8}$. Subsequently, the dishonest participant $P_{1}$ cannot get any idea about the marked state $|m\rangle$ from his operations, even though $D$ announces $S_{9}$. So, $P_{1}$ can acquire $D$'s secret shares with probability $\frac{9}{16}$. However, as previously mention, if $P_{1}$ resend two qubits of the triplet state to the participants $P_{2}$ and $P_{3}$ then $D$ can detect such deception with probability $\frac{12}{16}$. Moreover, assume that $D$ prepares any feasible $|S_{k}\rangle_{m}$, where $k=1,2,...,64$ and $m$ can be $000,001,010,011,100,101,110$ and $111$ with equal probability. The dishonest participant $P_{1}$ intercepts the three qubits and measure them in the basis $(9)$. So, $P_{1}$ gains no information about the marked state.

	\subsubsection{Entangle-Measure attack}
	\label{EM}
	The dishonest participant performs a more complex entangled-measure attack\cite{41,43}. The dishonest participant $P_{1}$ prepares an ancillary state $|0\rangle_{p}$ and that gets entangled with his qubit. $D$ is assumed to prepare $|S_{1}\rangle_{m=110}$ and $P_{1}$ entangles the auxiliary qubit with the sent qubit using a CNOT gate, then the combined quantum state becomes 
	\begin{align}
		|S_{1}\rangle_{m=110}\otimes |0\rangle_{p} \rightarrow |S_{1}\rangle_{m=110,p} = \dfrac{1}{\sqrt{2}}\Big(|0 \rangle |++\rangle|0\rangle_{p}+|1 \rangle(|++\rangle-|10\rangle)|1\rangle_{p}\Big).
	\end{align}
	Now suppose that $P_{1}$ does nothing and all participants $P_{1},P_{2}$ and $P_{3}$ operates the collective operation $U_{S_{1}MS_{1}}$ together. First they operate $U_{S_{1}}$ and gets
	\begin{align}\nonumber
		(U_{S_{1}}\otimes I)(|S_{1}\rangle_{m=110,p}) & = 
		\dfrac{1}{4\sqrt{2}}\Big[2(|100\rangle+|101\rangle+|110\rangle+|111\rangle)|0\rangle_{p}
		+(|000\rangle +|001\rangle \\ & \quad\quad\
		+|010\rangle+|011\rangle-|100\rangle-|101\rangle+3|110\rangle
		-|111\rangle)|1\rangle_{p}\Big],
	\end{align}
	from $(12)$, the participants get the marked state $|M\rangle$ i.e $|M\rangle=|110\rangle$ and they perform the operators $U_{M}$ and $U_{S_{1}}$ on $(12)$ respectively.
	\begin{align}\nonumber
		(U_{S_{1}MS_{1}}\otimes I) & (|S_{1}\rangle_  {m=110,p})	=
		\dfrac{1}{4\sqrt{2}}\Big[2(|100\rangle+|101\rangle+|111\rangle-|110\rangle)|0\rangle_{p}
		+(|000\rangle \\ & \quad\quad\quad
		+|001\rangle+|010\rangle+|011\rangle-|100\rangle-|101\rangle-|111\rangle-3|110\rangle)|1\rangle_{p}\Big].
	\end{align}
	Thus, $D$ can find $P_{1}$'s entanglement with probability $\frac{5}{32}$. Moreover, the honest participant detects such deception by observing the measurement results of the states $(12)$ and $(13)$. Instead, for any $(U_{S_{k}MS_{k}}\otimes I), ~k=1,2,...64$ the outcomes of $(U_{S_{k}MS_{K}}\otimes I)(|S_{1}\rangle_{m=110,p})$ is either  $|000\rangle,|001\rangle,|010\rangle,|100\rangle$ or  $|111\rangle$ with probability close to $\frac{5}{32}$. That is for any operation $(U_{S_{k}MS_{k}}\otimes I), ~k=1,2,...64$, the entanglement was detected by calculating the error in measurement results. Hence, the eavesdropper cannot perform successfully the entangle-measure attack  without error.

	\section{Comparative analysis}
	\label{CA}
	Most of the quantum secret sharing (QSS) schemes rely upon either entangled states or quantum Fourier transform (QFT) to secret sharing and secret reconstruction. The QSS schemes based on Grover's algorithm are introduced in \cite{30,31,32}. The proposed protocols based on Grover's algorithm can be regarded as encoding and decoding the secret by the same unitary operations. The work presented by Hsu in \cite{30} asserts a secure QSS scheme based on the idea of a two-qubit Grover's algorithm. In this scheme, the dealer (say Alice) encodes the single classical bits information into the two-particle quantum state and sends the qubits of this state to both participants (say, Bob and Charlie). After operating the collective operation, Bob and Charlie discuss their outcomes are perfectly correlated or anti-perfectly correlated. From here, they get the idea about the cheat-detecting state preparation of Alice. Whereas in our protocol, the dealer encodes three classical bits of information, and after operating the collective operation, the participants have no idea about the dealer's cheat-detecting state preparation. Moreover, based on initial state preparations, our scheme is $4$ times more secure than Hsu's scheme\cite{30}. In contrast to the QSS schemes in \cite{30,31,32}, our protocol has higher encoding capacity and security. The comparative analysis of our proposed scheme with some other QSS schemes based on the QSA is shown in Table \ref{tab.3}. 
	\begin{table}[htbp!]
		\caption{Comparision with schemes based on the QSA}
		\label{tab.3}
		\label{aggiungi}\centering \small
		\begin{tabular}{lccccc}
			\midrule
			Parameters  &  Hsu\cite{30} &  Hao et al.\cite{31} &  Tseng\cite{32} &  Proposed \\ \toprule
			Quantum resource & 2-qubit state &  2-qubit state & 2-qubit state &  3-qubit state \\
			Initial states & 16 & 16 & 4 & 64 \\
			Measurement bases & 1 & 3 & 1 & 1\\
			No. of unitary operations & 2 & 2 & 2 & 4\\
			Simulation analysis & no & no & no & yes\\ \midrule
		\end{tabular}
	\end{table}
	
	\section{Conclusion}
	\label{CO}
	The proposed scheme used Grover's three-particle quantum state to design a novel four-party secret sharing scheme. Any eavesdropping on an encoded state is identified. Moreover, the proposed protocol has been shown to be secure against Intercept attack, Intercept-Resend attack, and Entangle-Measure attack. The proposed scheme is simulated on the IBM quantum simulator. The current scheme has the scope to be applicable in quantum image secret sharing and generalization to a multiparty quantum secret sharing scheme.

	\paragraph{Acknowledgements} \small
	The first author with grant number 09/143(0951)/2019-EMR-I is grateful to Council of Scientific and Industrial Research (CSIR), India, for providing financial support to carry out this work. The work is also supported by SERB core grant CRG/2020/002040. The authors would like to thank to IBM for providing access to their Quantum-Experience cloud servers.



\begin{thebibliography}{99}
		
		\bibitem {1} Bennett, C.H.P., Brassard, G.: Quantum cryptography: public key distribution and coin tossing. Theor. Comput. Sci. \textbf{560}(12), 7–11 (2014)	
		
		\bibitem {2} Ekert, A.K.: Quantum cryptography based on bell’s theorem. Phys. Rev. Lett. \textbf{67}(6), 661 (1991)
		
		
		\bibitem {3} Bennett, C.H.: Quantum cryptography using any two nonorthogonal states. Phys. Rev. Lett. \textbf{68}(21), 3121 (1992)
		
		
		\bibitem{f} Long, G. L., Liu, X. S.: Theoretically efficient high-capacity quantum-key-distribution scheme. Phys. Rev. A \textbf{65}(3), 032302 (2002)
		
		\bibitem {4} Boström, K., Felbinger, T.: Deterministic secure direct communication using entanglement. Phys. Rev. Lett. \textbf{89}(18), 187902 (2002)
		
		
		\bibitem{5} Deng, F.G., Long, G.L., Liu, X.S.: Two-step quantum direct communication protocol using the einstein-podolsky-rosen pair block. Phys. Rev. A \textbf{68}(4), 042317 (2003)
		
		\bibitem{r} Xu, J.S., Li, C.F., Guo, G.C.: Silicon carbide based quantum networking. Fundamental Research 220-222 (2021)
		
		\bibitem{s} Kwek, L.C., Cao, L., Luo, W.,Wang Y., Sun, S., Wang, X., Liu, A.Q.: Chip-based quantum key distribution. AAPPS Bulletin \textbf{31}(15), (2021) 
		
		\bibitem{t} Qi, Z., Li, Y., Huang, Y., Feng, J., Zheng, Y., Chen, X.: A 15-user quantum secure direct communication network. Light: Science \& Applications \textbf{10}(183), (2021)
		
		
		\bibitem{6} Dušek, M.,Haderka, O., Hendrych, M.,Myška, R.: Quantum identification system. Phys. Rev.A \textbf{60}(1), 149 (1999)
		
		
		\bibitem{7} Zeng, G., Keitel, C.H.: Arbitrated quantum-signature scheme. Phys. Rev. A \textbf{65}(4), 042312 (2002)
		
		
		\bibitem{8} Lee, H., Hong, C., Kim, H., Lim, J., Yang, H.J.: Arbitrated quantum signature scheme with message recovery. Phys. Lett. A \textbf{321}(5–6), 295–300 (2004)
		
		
		\bibitem{9}  Li, Q., Chan,W.H., Long, D.-Y.: Arbitrated quantum signature scheme using bell states. Phys. Rev. A \textbf{79}(5), 054307 (2009)
		
		\bibitem{A} Zhou, R.G., Qian, W., Zhang, M.Q., Shen, C.Y.: Quantum image encryption and decryption algorithms based on quantum image geometric transformations. Int. J. Theor. Phys. \textbf{52}(6), 1802–1817 (2013)
		
		
		\bibitem{B} Liang, H., Tao, X., Zhou, N.: Quantum image encryption based on generalized affine transform and logistic map. Quantum Inf. Process. \textbf{15}(7), 2701–2724 (2016)
		
		
		\bibitem{C} Musanna, F., Kumar, S.: Image encryption using quantum 3-D Baker map and
		generalized gray code coupled with fractional Chen’s chaotic system. Quantum Inf. Process. \textbf{19}, 220 (2020)
		
		
		
		
		\bibitem{10} Shamir, A.: How to share a secret. Commun. ACM \textbf{22}(11), 612–613 (1979)
		
		
		\bibitem{11} Hillery, M., Bužek, V., Berthiaume, A.: Quantum secret sharing. Phys. Rev. A \textbf{59}(3), 1829 (1999)
		
		
		\bibitem{12} Cleve, R., Gottesman, D., Lo, H.K.: How to share a quantum secret. Phys. Rev. Lett. \textbf{83}(3), 648 (1999).
		
		
		\bibitem{13} Karlsson, A., Koashi, M., Imoto, N.: Quantum entanglement for secret sharing and secret splitting.Phys. Rev. A \textbf{59}(1), 162 (1999)
		
		
		\bibitem{14} Gottesman, D.: Theory of quantum secret sharing.Phys. Rev. A \textbf{61}(4), 042311 (2000)
		
		
		
		\bibitem{15} Guo, G.P., Guo, G.C.: Quantum secret sharing without entanglement. Phys. Rev. A \textbf{310}(4), 247–251 (2003)
		
		
		\bibitem{16} Xiao, L., Long, G.L., Deng, F.G., Pan, J.W.: Efficient multiparty quantum-secret-sharing schemes. Phys. Rev. A \textbf{69}(5), 052307 (2004)
		
		\bibitem{17} Li, Y., Zhang, K., and Peng, K.: Multiparty secret sharing of quantum information based on entanglement swapping. Phys. Lett. A \textbf{324}(5-6), 420–424 (2004)
		
		
		\bibitem{18} Deng, F.G., Long, G.L., Zhou, H.Y.: An efficient quantum secret sharing scheme with Einstein-Podolsky-Rosen pairs. Phys. Lett. A \textbf{340}(1–4), 43–50 (2005)
		
		
		\bibitem{19} Hsu, L.Y., Li, C.M.: Quantum secret sharing using product states. Phys. Rev. A \textbf{71}(2), 022321 (2005)
		
		
		\bibitem{20} Yan, F.L., Gao, T.: Quantum secret sharing between multiparty and multiparty without entanglement. Phys. Rev. A \textbf{72}(1), 012304 (2005)
		
		
		\bibitem{21} Markham, D., Sanders, B.C.: Graph states for quantum secret sharing. Phys. Rev. A \textbf{78}(4), 042309 (2008)
		
		
		\bibitem{22} Jia, H.Y., Wen, Q.Y., Qin, S.J., Guo, F.Z.: Dynamic quantum
		secret sharing. Phys. Lett. A \textbf{376}(10-11), 1034-1041 (2012)
		
		
		\bibitem{23} Yang, W., Huang, L., Shi, R., He, L.: Secret sharing based on quantum fourier transform. Quant. Inform. Process. \textbf{12}(7), 2465–2474 (2013)
		
		
		\bibitem{24} Maitra, A., De, S.J., Paul, G., Pal, A.K.: Proposal for quantum rational secret sharing. Phys. Rev. A \textbf{92}(2), 022305 (2015)
		
		
		\bibitem{p} Musanna, F., Kumar, S.: A novel three-party quantum secret sharing scheme based on Bell state sequential measurements with application in quantum image sharing. Quant. Inform. Process. \textbf{19}, 348 (2020)
		
		
		\bibitem{q} Musanna, F., Kumar, S.: Quantum secret sharing using GHZ state qubit positioning and selective qubits strategy for secret reconstruction. arXiv:2002.09182, (2020)
		
		
		
		
		\bibitem{25} Qin, H., Tso, R., Dai,Y.:Multi-dimensional quantum state sharing based on quantum Fourier transform. Quant. Inform. Process. \textbf{17}(3), 48 (2018)
		
		
		
		\bibitem{26}  Mashhadi, S.: General secret sharing based on quantum fourier transform. Quant. Inform. Process. \textbf{18}(4), 114 (2019)
		
		
		\bibitem{27} Sutradhar, K., Om, H.:  Efficient quantum secret sharing without a trusted player. Quntum Inf. Process. \textbf{19}(73), (2020)
		
		
		\bibitem{28} Jiang, S., Liu, Z., Lou, X., Fan, ., Wang, S., Shi, J.: Efficient verifiable quantum secret sharing schemes via eight quantum entangled states. International Journal of Theor. Phy. \textbf{60}, 1757-1766 (2021)
		
		
		
		\bibitem{29} Grover, L.K.: Quantum mechanics helps in searching for a needle in a haystack. Phys. Rev. Lett. \textbf{79}(2), 325 (1997)
		
		
		\bibitem{30} Hsu, L.Y.: Quantum secret-sharing protocol based on grover’s algorithm. Phys. Rev. A \textbf{68}(2), 022306 (2003)
		
		
		\bibitem{31} Hao, L., Li, J.L., Long, G.L.: Eavesdropping in a quantum secret sharing protocol based on Grover algorithm and its solution. Sci. China, Phys. Mech. Astron. \textbf{53}(3), 491-495 (2010)
		
		
		\bibitem{32} Tseng, H.Y., Tsai, C.W., Hwang, T., Li, C.M.: Quantum secret sharing based on quantum search algorithm. Int. J. Theor. Phys. \textbf{51}(10), 3101–3108 (2012)
		
		
		\bibitem{33} Wang, C., Hao, L., Song, S.Y., Long, G.L.: Quantum direct communication based on quantum search algorithm. Int. J. Quantum Inf. \textbf{8}, 443–450 (2010)
		
		
		\bibitem{34} Tseng, H.Y., Tsai, C.W., Hwang, T.: Controlled deterministic secure quantum communication based on quantum search algorithm. Int. J. Theor. Phys. \textbf{51}, 2447–2454 (2012)
		
		
		\bibitem{35} Zhang, W.W., Li, D., Song, T.T., Li, Y.B.: Quantum private comparison based on quantum search algorithm. Int. J. Theor. Phys. \textbf{52}, 1466–1473 (2013)
		
		
		
		
		\bibitem{36} Cao, H., Ma, W.P.: Multiparty quantum key agreement based on quantum search algorithm. Scientific reports. \textbf{7}, 45046 (2017)
		
		
		
		\bibitem{37} Huang, X., Zhang, S.B., Chang, Y., Qiu, C., Liu, D.M., and Hou, M.: Quantum key agreement protocol based on quantum search algorithm. Int. J. Theor. Phys. \textbf{60}(6), 1-10 (2021)
		
		
		\bibitem{38} Hao, L., Wang, C., Long, G.L.: Quantum secret sharing protocol with four state Grover algorithm and its proof-of-principle experimental demonstration. Optics Communication \textbf{284}(14), 3639-3642 (2011)
		
		\bibitem{39}  Vandersypen, L.M.K., Steffen, M., Sherwood, M.H., Yannoni, C.S., Breyta, G., Chuang, I.L.: Implementation of a three-quantum-bit search algorithm. Appl. Phys. Lett. \textbf{76}, 646 (2000)
		
		
		\bibitem{40} IBM quantum computing platform. https://www.ibm.com/quantum-computing/
		
		
		\bibitem{41}  Ekert, A.K., Huttner, B., Palma, G.M., Peres, A.: Eavesdropping on quantum-cryptographical systems. Phys. Rev. A \textbf{50}, 1047 (1994) 
		
		
		\bibitem{42}  Biham, E., Mor, T.: Security of quantum cryptography against collective attacks. Phys. Rev. A \textbf{78}, 2256 (1997) 
		
		
		\bibitem{43} Cirac, J.I., Gisin, N.: Coherent eavesdropping strategies for the four state quantum cryptography protocol. Phys. Lett. A \textbf{229}, 1 (1997) 
		
		
		
		
		
	\end{thebibliography}
\end{document}